\documentstyle[aps]{revtex}

\begin{document}
\title{Entangled two atoms through different couplings and the thermal noise }
\author{L. Zhou, X. X. Yi, H. S. Song and Y. Q. Guo.}
\address{{\small Department of Physics, Dalian University of Technology,}\\
Dalian,116024, P. R. China.}
\maketitle

\begin{abstract}
The entanglement of two atoms is studied when the two atoms are coupled to a
single-mode thermal field with different couplings. The different couplings
of two atoms are in favor of entanglement preparation: it not only makes the
case of absence entanglement with same coupling appear entanglement, but
also enhances the entanglement with the increasing of the relative
difference of two couplings. We also show that the diversity of coupling can
improved the critical temperature. If the optical cavity is leaky during the
time evolution, the dissipative thermal environment is benefit to produce
the entanglement.

\end{abstract}

\section{Introduction}

Environment noise can result in decoherence of quantum system. In order to
overcome the decoherence, some authors proposed a number of approaches such
as a quantum error correction\cite{quco} and a decoherence-free subspace\cite
{dfre}. Another direct method is to employ the environment noise to produce
entanglement [3-12]. In solid state system, the thermal entanglement in
Heisenberg model is a example of using the thermal environment [3-6]. It has
been found that the anisotropy in XY Heisenberg chain can be used to
increase the critical temperature\cite{GL}. In cavity QED, the cavity loss
or the thermal field can also induce entangment between two atoms [7-11]. In
Ref.\cite{plenio}, using quantum jump approach, the authors described a
scenario where entanglement between two atomic systems can be induced via
continous observation of the cavity loss. Working in different initinal
state (different from Ref. \cite{plenio}) and employing a master equation,
in Ref. \cite{xxy}, the author proposed a sheme to prepare two atomic
entanglement. Two two-level atoms can be entangled through interacting with
a single-mode thermal field \cite{kim} \cite{zl}. But in the studies on
cavity QED\cite{bose }\cite{kim} \cite{zl} , for simplicity of the
calculation , the couplings $g_{_i}(i=1,2)$ between the two atoms and the
field are assumed equal. Although, in Ref.\cite{plenio}\cite{xxy}, their
calculation also applied to $g_{_1}\neq g_{_2}$, they just had discussed the
equal couplings in their plots and had not discussed the effect of two atoms
with different couplings. In fact, the coupling rate $g$ depends on the atom
position ${\bf r}$ ; due to the randomness of the atom position ${\bf r}$,
it is very difficult to control the same couplings between different atom 
\cite{duan}. The truth that the anisotropy in two-qubit Heisenberg model can
increase the critical temperature motivate us to consider whether different
couplings $g$ of the two atoms result in some novel properties in cavity QED.

In this paper, we study the system that the two atoms with different
coupling constants are coupled to the optical cavity which is initinally in
thermal field state. When the atoms are initially in $|ee\rangle $ and the
cavity is initially in a single-mode thermal field, with the same coupling
constant they are not entangled \cite{kim} but if with different coupling
constant they can be entangled. We analyse the dependence of the relative
difference of two couplings on the entanglement. We find that there is a
possiblity to obtian more entanglement through increasing the relative
difference of two couplings. We show that the diversity of coupling can
improved the critical temperature at which entanglement disappears. We also
study the case of a cavity with a steady leak during the time evolution, it
is found that the dissipative cavity field is benefit to produce the
entanglement.\smallskip

\section{The effect of the diversity coupling of the two-atom}

In order to make clear the function of different couplings, we first
consider a quantum system composed of two two-level atoms interacting with a
single-mode thermal field, which can be produced by a leaky cavity in
thermal equilibrium at temperature $T$. After preparation the initial cavity
field, the cavity stop to leak. The cavity mode is assumed to be resonant
with the atomic transition frequency. Under the rotating wave approximation,
the Hamiltonian in the interacting picture is 
\begin{equation}
H_I=g_{_1}(a\sigma _1^{+}+a^{+}\sigma _1^{-})+g_{_2}(a\sigma
_2^{+}+a^{+}\sigma _2^{-})
\end{equation}
where $a$ $(a^{+})$ denotes annihilation (creation) operator of the cavity
field; the atomic transition operators are $\sigma _i^{-}=|g\rangle
_i\langle e|$ and $\sigma _i^{+}=|e\rangle _i\langle g|;$ $g_{_1}$ and $%
g_{_2}$ are the two coupling constants for the atom $i$ $(i=1,2)$. For
expression the diversity of the two couplings, we let 
\begin{equation}
g=\frac{g_{_1}+g_{_2}}2,\text{ }\gamma =\frac{g_{_1}-g_{_2}}{g_{_1}+g_{_2}}
\end{equation}
where $g$ denote average coupling and $\gamma $ express the relative
difference of the two atomic couplings. Because of the identity of the two
atoms, the range of $\gamma $ is between $0$ and $1$. The Hamiltonian can be
rewritten as 
\begin{equation}
H_I=g(1+\gamma )(a\sigma _1^{+}+a^{+}\sigma _1^{-})+g(1-\gamma )(a\sigma
_2^{+}+a^{+}\sigma _2^{-})
\end{equation}
The single-mode thermal field with its mean photon number $\bar{n}$ is a
mixture of Fock states. It take the form

\begin{equation}
\rho _{c}=\sum_{n}\frac{\bar{n}^{n}}{(1+\bar{n})^{n+1}}|n\rangle \langle n|,
\end{equation}
and $\bar{n}=(e^{\frac{\hbar \omega }{k_{B}T}}-1)^{-1}$ relates to the
environment temperature. Entanglement of two atoms can be measured by
concurrence $C$ which is written as\cite{chb}\cite{sh}

\begin{equation}
C=\max (0,2\max \{\lambda _i\}-\sum_{i=1}^4\lambda _i),
\end{equation}
where $\lambda _i$ is the square roots of the eigenvalues of the matrix $%
R=\rho S\rho ^{*}S$, $\rho $ is the density matrix, $S=\sigma _1^y\otimes
\sigma _2^y$ and $*$ stands for complex conjugate. The concurrence is
available no matter what $\rho $ is pure or mixed. We will numerical
calculate the entanglement between the two atoms.

In Ref. \cite{kim}\cite{zl}, for simplicity they assumed the atoms 1 and 2
couple to a single-mode thermal field with the same coupling constant. They
found that if the two atoms initially are both in excited states, the two
atoms could not be entangled neither in one-photon process nor in two-photon
process. In our simulation, if $\gamma =0,$ we also can not find
entanglement. But when we chose $\gamma =0.4$ , the result is reversed which
is shown in Fig. 1. It is obvious that under the condition of different
couplings the two atoms (initial in $|ee\rangle $ ) can be entangled. This
is contrary to the case of the same coupling constants\cite{kim}\cite{zl}.
When the two atoms initially are both in excited state, it had been shown
that the atom and the field are always entangled despite of the presence of
the other atom\cite{kim}. It is the atom-field entanglement that result in
the decoherence between the two atoms. We can understand it from the same
couplings: becasuse their couplings are the same, their rabi oscillation
also have the same steps, so the two atoms have no correlation, i.e., there
is no entanglement between them. If the couplings are not the same, when
they resonate with the cavity, they are no longer with the same step.
Through interacting with cavity, the two atoms are entangled. As we state
above, the different couplings are more closely to experiments. So, the
existing entanglement with different couplings maybe more easy achieved.
Fig. 1 also show that the vacuum state ( $\bar{n}=0$) induces the maximum
entanglement. With the increase of the mean photon number, the entanglement
between two atoms decreases. At a certain value of $\bar{n}$, the
entanglement disappears; we call that $\bar{n}$ as critical average photon
number and the temperature ( corresponding to the $\bar{n})$ is called as
critical temperature.

Increasing $\gamma $ to $0.8$, we plot the entanglement of two-atom in
Fig.1b. Comparing Fig.1a with Fig.1b, one can find that the maximum
entanglement of Fig. 1b is larger than that of Fig. 1a, that is, one can
obtain much more strong entanglement with $\gamma =0.8$. On the other hand,
in Fig. 1a the critical average photon number is below 4 and in Fig.1b when $%
\bar{n}=5$, the maximum entanglement is about 0.15, that is, the critical
average photon number is larger than 5. So, the critical temperature of Fig.
1b is larger than that of Fig.1a. Whether can we conclude from Fig.1 that
the larger the relative difference of two couplings, the larger the critical
average photon number, the larger the critical temperature? The answer is
no. Because the rabi oscillation of the atoms is relevant to the couplings,
the entanglement is not a monotonic function of $\gamma $ . It will be shown
in Fig. 2 and we will analyse it laterly. But at least in some ranges we can
say that through increasing $\gamma $, the critical temperature can be
improved. The conclusion, in a certain extent, is similar to that the
anisotropy in Heisenberg XY chain can increase critical temperature of the
thermal entanglement \cite{GL}. The difference is that there, the critical
temperature of the thermal entanglement is monotonously increased with the
increasing of the anisotropy parameter, here due to the rabi oscillation,
the critical temperature is not a monotonous function of the two couplings.
But with the increasing of $\gamma $ , there are the possiblity and the
tendency to increasing the critical temperature (we will explain it next).

Because every atom periodically entangles and disentangles with the cavity
field, and the periodicity relates to the coupling, so the entanglement
between two atoms also exhibit periodicity and its periods are also related
to the two couplings. We can observe this periodicity in Fig.1. For make
clear the dependence of the entanglement on $\gamma ,$ we plot the
entanglement as a function of $\gamma $ and $t$ in Fig. 2 for the atomic
initial state $|ee\rangle $, where the average photon number $\bar{n}=1$. We
can see that for $\gamma =0$ no entanglement exist which is coincident with
the Ref.\cite{kim}. But with the increasing of $\gamma $, in som time
intervals one can see the large entanglement. Of cause, due to existing many
peaks (existing up the hill and down the hill) , there are either the extent
in which entanglement is increased with the increasing of $\gamma $ or the
extent in which the entanglement is decreased with the increasing of $\gamma
.$ Although the relation between the entanglement and $\gamma $ is not
monotonous, the main trend of entanglement is increased with the increasing
of $\gamma $. In other word, the maxima values of entanglement is increased
with the increasing of $\gamma .$ Thus, if we can keep the relative large $%
\gamma $, even the two atoms interact with thermal field, we can also obtain
relative large entanglement. On the other hand, since the average photon
number is a measure of the cavity classicality, the larger it is, the
smaller the entanglement, i.e., with the increase of the mean photon number,
the entanglement gradually decreases to zero. If in the extent in which
entanglement is increased with the increasing of $\gamma ,$ we can reckon
the critical average photon number is increased with the increasing of $%
\gamma $, thus the critical temperature is increased. If the span of $\gamma 
$ is relative large such as $\gamma $ from 0.4 to 0.8, the maxima
entanglement is increased, the critical temperature, of cause, is increased.

The two atoms with initial state $|ee\rangle $, which can not be entangled
if they have the same coupling constants \cite{kim}\cite{zl}, can be
entangled when they are different in coupling constant. The initial state $%
|eg\rangle $ is the best case which the entanglemet could be the relative
best but is far smaller than 1\cite{kim}\cite{zl}. If the coupling constants
are different, what will happen ? We directly numerical simulate the
entanglement as a function of $\gamma $ and $t$ in Fig. 3. In Fig. 2 and
Fig. 3, we both can see that when $\gamma =1$ the entanglement do not exist.
For $\gamma =1$, only the one atom interact with the cavity and the other
atom has no coupling with the field, so, there is no entanglement between
them. When the two-atom is initially in the state $|eg\rangle $ with $\gamma
=0$ is exact the case which had been discussed in Ref.\cite{kim}. From Fig.
3, we can observe entanglement when $\gamma =0,$ and it is different from
the case of initial state $|ee\rangle $ with $\gamma =0$. But the
entanglement in Fig. 3 when $\gamma =0$ is not the largest. With the
increasing of $\gamma ,$ the maximum values of entanglement are also
increased. There are still existing many ''hills'', that is, there are
either the extent in which entanglement is increased with the increasing of $%
\gamma $ or the extent in which the entanglement is decreased with the
increasing of $\gamma .$ Therefore, we can also obtain strong entanglement
and improve the critical temperature through increasing $\gamma $ in some
extent. Thus, although Fig. 3 is obvirously different from Fig. 2, the
effects of $\gamma $ on increasing critical temperature and improved
entanglement are exactly the same in these two kinds of initial states.

\section{The influence of thermal noise on the atom-atom entanglement}

When the single-mode cavity is in thermal equilibrium with its environment
due to the leakage, the cavity is in thermal field \cite{gardiner}. In
section ${\it 2}$ and in Ref.\cite{kim}\cite{zl}, the thermal field is just
as initial cavity state; in the process of evolution, the cavity stop
leaking. Since the cavity is leaky, why we assumed it stop to leaking? Now,
we let it continue to leak in the later time evolution. \ 

The master equation governing the time evolution of the global system is
given by ($\hbar =1$) 
\begin{equation}
\dot{\rho}=i[\rho ,H]+{\cal L}(\rho ),
\end{equation}
where the Hamiltonian still have the form of Eq.(3). The Liouvillean is
given by 
\begin{equation}
{\cal L}(\rho )=\kappa (\bar{n}+1)(2a\rho a^{+}-a^{+}a\rho -\rho
a^{+}a)+\kappa \bar{n}(2a^{+}\rho a-aa^{+}\rho -\rho aa^{+}
\end{equation}
We chose $\kappa =0.4$, $\gamma =0.4$ and simulate the case of which the two
atoms are initially in $|ee\rangle $ and the cavity is initially in the
thermal state. Figure 4 show the entanglement as a function of average
photon number and time $t$. We notice that the critical average photon
number is larger than 5, and in Fig. 1 $\bar{n}$ is smaller than 5. To a
thermal field state, the dissipative environment can increase the critical
average photon number. Furthermore, \smallskip with the time evolution, the
entanglement is no longer periodic appearing zero but gradually reaches
their asymptotic values, and the asymptotic values decrease with the
increasing of $\bar{n}$. When the cavity field initially is in thermal
state, the dissipation of the cavity virtually can be considered as a driver
of the \smallskip cavity, so, with the evolution the entanglement is
gradually increased. This is contrary to the general case in which the
entanglement is destroyed by environment due to the decoherence. After all
the system is a open one, so the entanglement is no longer periodic
appearing zero but gradually reaches their asymptotic values. So, after long
time evolution, we can obtain a strong and steady entanglement. Therefore,
as long as the evolution time is enough, it is not necessary to precise
control the interaction time. In experiment, precise control of the
interacting time is very difficult to achieve \cite{browne}. We summarize
that the dissipative environment not only improved th critical temperature
but also provide us a steady entanglement.

\section{Conclusion}

We discuss the entanglement induced by a single-mode heat environment when
the two atoms are coupled to the optical cavity with different coupling
constants. When the atoms are initially in $|ee\rangle $, with the same
coupling constant they could not be entangled \cite{kim} but if with
different coupling constant they can be entangled. Even to the initial state 
$|eg\rangle $ in which the two atoms can be entangled with the same
coupling, the different couplings are avail to produce atom-atom
entanglement. Through the analysis about the dependence of entanglement on
the relative difference of two couplings $\gamma $, we find that by
increasing $\gamma $ we can obtain strong entanglement. We also show that
the diversity of coupling can improved the critical temperature, which is
very similar to that the anisotropy in Heisenberg XY chain can increase the
critical temperature of thermal entanglement \cite{GL}. We study the leak of
the cavity on the entanglement during the time evolution. It is found that
the keeping on leaking is benefit to produce the entanglement. \smallskip It
not only can improves the critical temperature but also provid us a relative
steady and strong entangled state.

Figure captions:

Fig.1. Entanglement as a function of average photon number $\bar{n}$ and $t$
when the pair of atoms is initially prepared in the state $|ee\rangle $%
\smallskip and the field is initially in the thermal state, where (a) : $%
\gamma =0.4,$ (b):$\smallskip \gamma =0.8$; for all plots $g=1$.

Fig. 2. The dependence of entanglement on the relative difference of the two
atomic couplings $\gamma $ and time $t$ for the initial atomic state $%
|ee\rangle $ when $g=1$, $\bar{n}=1.$

Fig. 3. The same as Fig. 2 but for the initial atomic state $|eg\rangle .$

Fig. 4. The effect of dissipative environment on the atom-atom entanglement
when the atoms are initially in $|eg\rangle $ for $\kappa =0.4$, $g=1$.


\begin{references}
\bibitem{quco}  P. W. Shor, Phys. Rev. A 52, R2493 (1995); A.M. Steane,
Phys. Rev. Lett. 77, 793 (1996).

\bibitem{dfre}  D. A. Lidar, I. L. Chuang, and K.B. Whaley , Phys. Rev.
Lett. 81, 2594 (1998); A. Beige, D. Braun, B. Tregenna, and P. L. Knight,
Phys. Rev. Lett. 85, 1762 (2000).

\bibitem{wxg3}  X. Wang, Phys. Rev. A 64, 012313 (2001); X. Wang, Phys. Rev.
A 66,034302 (2002);X. Wang, Phys. Rev. A 66,044305 (2002).

\bibitem{MC}  M. C. Arnesen, S. Bose and V. Vedral, Phys. Rev. Lett. 87,
017901 (2001).

\bibitem{GL}  G. L. Kamta and A. F. Starace, Phys. Rev. Lett. 88,
107901(2002).

\bibitem{zl1}  L. Zhou, H. S. Song, Y. Q. Guo and C. Li, Phys. Rev. A 68,
024301 (2003)

\bibitem{plenio}  M. B. Plenio, S. F. Huelga, A. Beige, and P. L. Knight,
Phys. Rev. A 59, 2468 (1999); A. Beige, H. Cable, and P.L. Knight,
quant-ph/0303151.

\bibitem{bose }  S. Bose, I. Fuentes-Guridi, P. L. Knight, and V. Vedral,
Phys. Rev. Lett. 87, 050401 (2001)

\bibitem{kim}  M. S. Kim , Jinhyoung Lee, D. Ahn and P. L. Knight, Phys.
Rev. A 65, 040101(2002)

\bibitem{zl}  L. Zhou, H. S. Song, and C. Li, J. Opt. B: Quantum Semiclass.
Opt. 4, 425 (2002).

\bibitem{xxy}  X. X. Yi, C. S. Yu, L. Zhou, and H. S. Song,
quant-ph/0306091(to appear in Phys. Rev. A).

\bibitem{braun}  D. Braun, Phys. Rev. Lett. 89, 277901 (2002).

\bibitem{duan}  L. M. Duan, A. Kuzmich, and H. J. Kimble, Phys. Rev. A 67,
032305 (2003).

\bibitem{chb}  C. H. Bennett, D. P. DiVincenzo, J. A. Smolin and W. K.
Wootters, Phys. Rev. A 54, 3824 (1996)

\bibitem{sh}  S. Hill and W. K. Wootters, Phys. Rev. Lett. 78, 5022 (1997).

\bibitem{gardiner}  C. W. Gardiner and P. Zoller, Quantum Noise (spinger
2000).

\bibitem{browne}  D. E. Browne and M. B. Plenio, Phys. Rev. A 67, 012325
(2003).
\end{references}
\end{document}